\documentclass[%
reprint,
superscriptaddress,
amsmath,amssymb,
prstab,
]{revtex4-1}

\usepackage{placeins}
\usepackage{enumerate}
\usepackage{csquotes}
\usepackage{graphicx}
\usepackage[dvipsnames]{xcolor}
\usepackage{amsmath}
\usepackage{natbib}
\usepackage[nottoc,numbib]{tocbibind}
\usepackage[colorlinks]{hyperref}    
\graphicspath{ {./images/} }
\begin{document}

\preprint{AIP/123-QED}

\title{High-Power Ferro-Electric Fast Reactive Tuner}

\author{Ilan \surname{Ben-Zvi}}
\email{benzvi@bnl.gov}
\affiliation{CERN, CH-1211 Geneva, Switzerland}
\affiliation{Collider-Accelerator Dep., BNL, Upton NY USA}

\author{Alejandro \surname{Castilla}}
\affiliation{Engineering Department, Lancaster University, LA1 4YW, UK}
\affiliation{CERN, CH-1211 Geneva, Switzerland}

\author{Alick \surname{Macpherson}}
\email{alick.macpherson@cern.ch}
\affiliation{CERN, CH-1211 Geneva, Switzerland}

\author{Nicholas \surname{Shipman}}
\affiliation{CERN, CH-1211 Geneva, Switzerland}

\date{\today}

\begin{abstract}
We present a novel design of a FerroElectric Fast Reactive Tuner (FE-FRT) capable of modulating Mega VAR reactive power on a sub-microsecond time scale. We show detailed analytical estimates of the performance of this device and benchmark these estimates against finite element method eigenmode and frequency domain electromagnetic simulations.
\end{abstract}

\pacs{29.20.−c, 29.20.Ej, 41.75.Fr, 87.56.bd, 29.27.-a}
\keywords{ SRF, Cavity, RF, fast tuning}
\maketitle

\section{Introduction}\label{Introduction}
Fast tuning at high power is an enabling technology in particle accelerators. Uses are abundant and include: avoidance of beam resonances during acceleration in storage ring cavities; compensation of transient beam loading; and correction of microphonics in superconducting cavities.

Reactive tuning is the controlled change of the resonant frequency of a cavity by coupling the cavity to a variable reactance.  Its performance is characterized by the amount of variation in reactive power that the tuner delivers to the cavity as well as the resultant power dissipation in the tuner. There is a long history of the development of reactive tuning. As an example, the highest reactive power tuner was a VCX (voltage controlled reactance) developed at Argonne National Laboratory for the control of acoustic frequency noise of superconducting resonators in a linac. That device, using PIN diodes, achieved a peak reactive power of 20 kVAR with a 70 watt power dissipation in the diodes. \cite{Bogaty}

Ferroelectric materials provide a fast, low loss variable reactance for such an application. The first demonstration of frequency control of a cavity was made with a ferroelectric tuner based on a coaxial geometry in 2019 at CERN \cite{Shipman2019}.  More recent tests at CERN have measured the speed of the cavity frequency shift to be $<600\,\textrm{ns}$\cite{Shipman2021} many times faster than the time constant of the cavity and currently believed to be limited only by the speed of the external powering circuit.  The response time of bulk ferroelectric similar to the material used in the prototype FE-FRT has been measured at $<30ns$\cite{Kazakov2009}.

The objective of this paper is to present the design of a new class of FE-FRT and evaluate its performance. We will show an example in which this novel approach is capable of providing reactive power tuning in excess of $10^7$ VAR with a figure of merit greater than 200. 

\begin{figure}[htb]
\centering
\includegraphics[width=0.9\columnwidth]{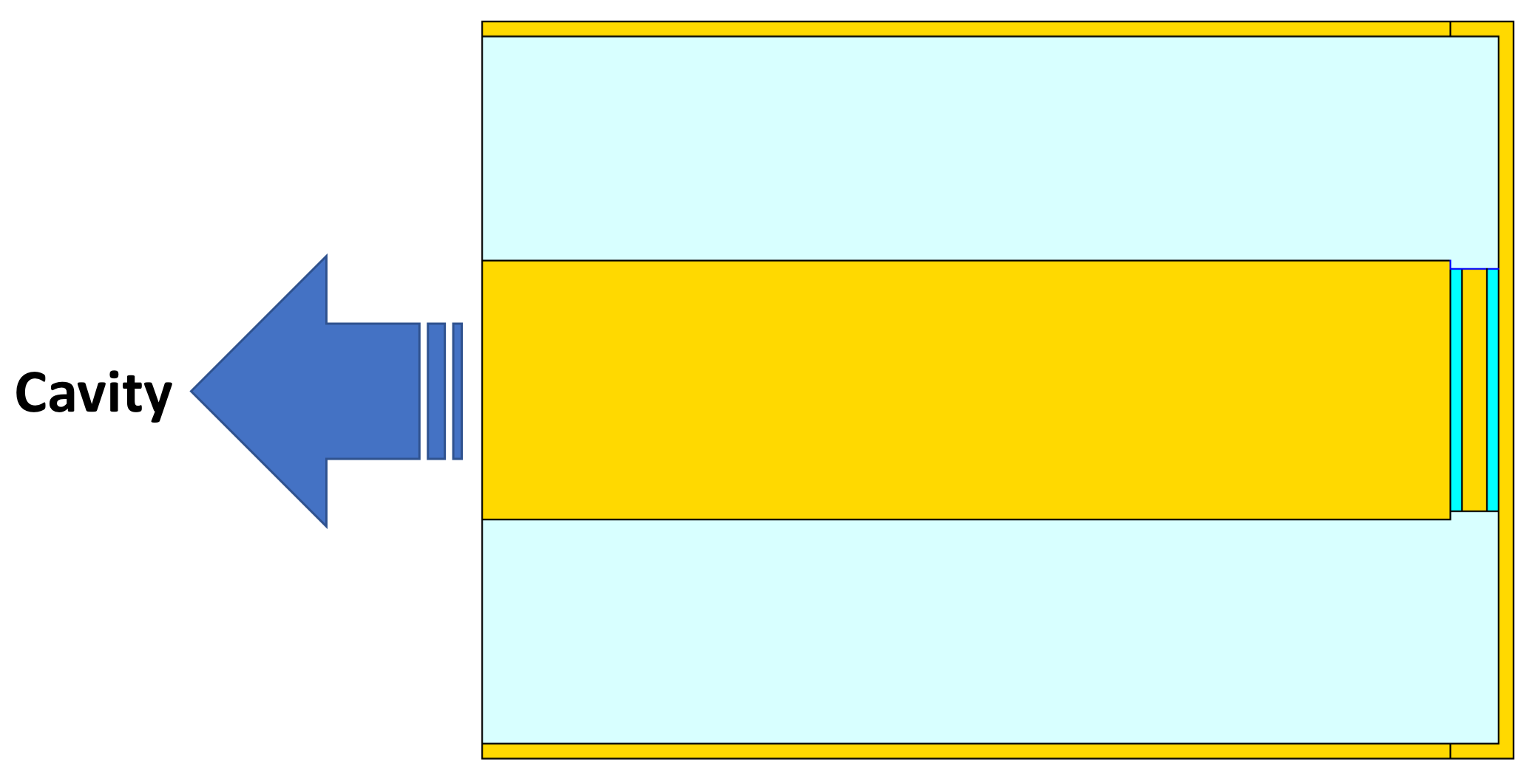}
\caption{\label{fig:Slide1} A schematic diagram of a simple two-wafer capacitively terminated, coaxial transmission line ferroelectric tuner. The ferroelectric wafers are shown in green, copper in yellow. Cooling and bias connections to the inter-wafer electrode are not shown in this figure.}
\end{figure}

The device we describe is a multi-layer parallel-plate capacitor coupled to the cavity via a short, coaxial transmission line as shown in fig.\ref{fig:Slide1}. The capacitor's dielectric material is ferroelectric. The dielectric constant of the ferroelectric is modulated by a voltage signal applied to electrodes located between the ferroelectric wafers.

Notable features of this design include: 

\begin{enumerate}[a.]
    \item a stack geometry where multiple capacitors made of ferroelectric wafers and metalic spacers are connected in series, where the spacers provide thermal and electrical contact to the wafers and the impedance of the stack can be optimized;
    \item the use of flat thin wafers of the ferroelectric material, thus providing a high modulation electric field for a given voltage and large heat transfer through the wafer;
    \item in built isolation of high voltage from the cavity wall and coupling antenna;
    \item the option of using a number of stacks to optimize the dimensions of the wafers and optionally to facilitate tailoring the temporal modulation of the tuner by allowing different voltage patterns to be applied to different stacks.
	
\end{enumerate}

We will start in Section \ref{FoM} by defining a highly useful figure of merit. Then in Section \ref{Analytic} we present an intuitive analytic model of a basic, simplified tuner device backed-up by precise numerical simulations, predicting Mega-VAR  reactive power capability. We will address the physical realization of the new ferroelectric tuner concept in Section \ref{Physical}, including the thermal load consideration, and the considerations of the application of the bias (modulation voltage). Then, in Section \ref{Basic} we will present a basic example of a stacked capacitor tuner and compare the analytic model with numerical simulations. Next, in Section \ref{Coupling} we will connect the tuner as developed in the previous sections to a cavity, represented by a discrete equivalent circuit and evaluate the frequency shift produced by the tuner, the requirements on the external coupler and the RF power load on the tuner. This analytic model will be backed up by numerical simulations. Finally, in Section \ref{Application} we will work out in details a relevant but challenging application of a high reactive-power FE FRT to an 80 MHz CERN PS cavity, and present the performance of such an FE-FRT tuner.

\section{The Figure of Merit}\label{FoM}

It is useful to quantify the performance of a variable reactance tuner by a Figure of Merit (FoM) representing a benefit to cost ratio. The benefit of the device is the change in reactive power. The cost function is the dissipated power in the tuner.   Specifically we define the FoM as:
\begin{eqnarray}\label{eq:FoM}
    \text{FoM} &=& \frac{\text{Change in reactive power}}{\text{Twice the average dissipated power}} \nonumber\\
    &=&  \frac{|\Delta\mathcal{P}_{if}|}{2\sqrt{P_iP_f}}
\end{eqnarray}
where $\Delta\mathcal{P}_{if}$ is the change in reactive power between the initial and final states and $P_i$ and $P_f$ are the powers dissipated in the FE-FRT in the initial and final states respectively. The factor two and geometric average of the dissipated powers are added to be consistent with previous literature.  The most relevant value for the denominator in eq.~\eqref{eq:FoM} would be a time weighted average of the dissipated power, which would depend on the temporal distribution of the required reactive power. For the model to be general however, we cannot assume any particular weighting function, fortunately for practical applications the dissipated power is often only weakly dependent on the applied voltage and is well represented as a geometric average of the end points  $P=\sqrt{P_iP_f}$. 

Defining the impedance of the FE-FRT as presented to the cavity as:
\begin{equation}\label{eq:Zfrt}
    Z = R + jX
\end{equation}
definitions for the reactive and dissipated power in the tuner can be easily derived as:
\begin{equation}\label{eq:Pdis}
    \mathcal{P} = I^2X
\end{equation}
and
\begin{equation}\label{eq:Preac}
    P = I^2R
\end{equation}
respectively, where $I$ is the current at the cavity FE-FRT transition.  Substituting eq.~\eqref{eq:Pdis} and eq.~\eqref{eq:Preac} into eq.~\eqref{eq:FoM} we obtain:
\begin{equation}\label{eq:FoMcircuit}
    \text{FoM} = \frac{|X_i-X_f|}{2\sqrt{R_iR_f}}
\end{equation}

Clearly, a differential form of eq.~\eqref{eq:FoMcircuit} also applies,

\begin{equation}\label{eq:FoMdiff}
    \dot{\text{FoM}} = \frac{|\dot{X}|}{2R}
\end{equation}

It is useful and instructive to consider a tuner consisting only of a ferroelectric capacitor connected directly to the cavity (or with a transmission line of vanishingly short length). Let us define the bare capacitance (without the ferroelectric material) as C and let the relative dielectric constant of the ferroelectric be $\varepsilon (1-j \tan\delta)$, where $\tan\delta$ is the loss tangent.

The impedance of the capacitor with a relative dielectric constant given above, at an angular frequency $\omega$ is  given by:

\begin{equation}	
    \text{Z}_C = \frac{-j}{\omega C \varepsilon (1-j\tan\delta)} 
    \label{eq:Zcapacitor}
\end{equation}

Writing the impedance in terms of real and imaginary components,

\begin{equation}\label{eq:Z-real-and-imaginary}	
    \text{Z}_C =  \frac{\tan\delta}{\omega C\varepsilon(1+\tan^2\delta)} -j\frac{1}{\omega C\varepsilon(1+\tan^2\delta)}
\end{equation}

 We now substitute eq.~\eqref{eq:Z-real-and-imaginary} into the definition for the FoM given in eq.~\eqref{eq:FoMcircuit} to obtain:
 \begin{equation}\label{eq:FoMcapacitor}
     \text{FoM}_C = \frac{\varepsilon_i-\varepsilon_f}{2\tan\delta\sqrt{\varepsilon_i\varepsilon_f}}
 \end{equation}

Where we have neglected terms in $\tan^2\delta$ compared to 1.  Where $\frac{\varepsilon_i}{\varepsilon_f}<2$, as is to be expected, eq.~\eqref{eq:FoMcapacitor} can be well approximated as:
\begin{equation}\label{eq:FoMcapacitor2}
     \text{FoM}_C \approx \frac{1}{2\tan\delta} \ln{\frac{\varepsilon_i}{\varepsilon_f}}
 \end{equation}
Note that the capacitor's figure of merit $\text{FoM}_C$ can be quite high. Given a dielectric constant ratio of 1.4 and a loss tangent of 0.001, $\text{FoM}_C$ reaches 168, leading to a ratio of the reactive to dissipative power of 336.  It is interesting to note that in this idealized case, the FoM does not depend on the absolute value of the capacitance.

As we will see below, the transmission line introduces dissipation in addition to the resistive power dissipation in the ferroelectric wafers, the FoM of the tuner will be lower than that of just the $\text{FoM}_{C}$ of the capacitor. 
Which figure of merit is relevant? That depends on the application. For a FE-FRT tuning with relatively low reactive power, of the order of $10\,\textrm{kVAR}$ or less, it is the complete device FoM that we should consider, since the power dissipation in the ferroelectric is easily handled and our aim is to reduce the power of the RF amplifier driving the cavity. The fast tuning of energy recovery linacs for microphonics compensation would be an example of this regime. On the other hand, when the reactive power demands are high, say of the order of MVAR, the challenge is the cooling of the ferroelectric wafers, and the dissipation in the coupler and transmission line is less of a concern. Normal conducting cavities, where the frequency has to be quickly tuned over a large range would be an example of this.

\section{Analytic model}\label{Analytic}

Let us take a closer look at an FE-FRT device, consisting of a dissipative transmission line terminated by a variable capacitor, where as before the capacitance is varied using a ferroelectric dielectric material tuned by an applied voltage. In section \ref{Coupling} we will couple the tuner device to a cavity.

The complex propagation coefficient of the transmission line is given by
\begin{equation}	
    \gamma = \alpha + j \beta
    \label{eq:TL}
\end{equation}

Where  $\beta =2 \pi/\lambda$, $\lambda$ is the wavelength of the resonator’s frequency, and $\alpha$ is the attenuation constant of the line. 
For a vacuum filled, slightly dissipative coaxial line with inner conductor of radius $a$, outer conductor of radius $b$ and surface resistance $R_s$ we get
\begin{equation}	
    \alpha =  (\frac{1}{a} + \frac{1}{b}) \frac{R_s}{2 \eta \log (b/a)}
    \label{eq:TL_attenuation}
\end{equation}

Where $\eta$ is the impedance of vacuum, with $\eta \approx 376.7 \Omega$. We assume a good line, $\alpha \ll \beta$. For a high-quality copper, the conductivity (room temperature) is $\sigma=6 \times 10^7 [\Omega m]^{-1}$. Neglecting the anomalous resistance: 
\begin{equation}	
    R_s = \sqrt{\frac{\omega\mu_0}{2 \sigma}}
    \label{eq:R_s}
\end{equation}

The impedance of a transmission line terminated by impedance $Z_L$ is given by
\begin{equation}	
    \dfrac{Z}{Z_0}= \frac{\frac{Z_L}{Z_0} +\tanh(\gamma l)}{1+\frac{Z_L}{Z_0} \tanh(\gamma l)}
    \label{eq:Z_definition}
\end{equation}

 $Z_0$ and $l$ are the characteristic impedance and length of the transmission line leading from the resonator’s port to the termination, and $Z_0$ is given by

\begin{equation}	
    Z_0= \frac{\eta}{2\pi}\log{\frac{b}{a}}  (1-j \frac{\alpha}{\beta})
    \label{eq:Z0_definition}
\end{equation}

Let us define the capacitance of the capacitor without the ferroelectric material as $C$ and let the relative dielectric constant of the ferroelectric be $\varepsilon$, and $\delta$ is the loss tangent.

Therefore, the terminating impedance is
\begin{equation}	
    	Z_L=\frac{1}{j \omega C \varepsilon (1-j\tan\delta)}
    \label{eq:ZL}
\end{equation}

 With the definition $ \zeta\equiv Z_L/Z_0 $  we obtain
\begin{equation}	
    Z = Z_0  \frac{\zeta+ \tanh(\gamma l)}{1+\zeta \tanh(\gamma l)}=R+jX
    \label{eq:Z}
\end{equation}

Now we can apply equation \ref{eq:Z} for the impedance Z to the evaluation of the FoM based on the differential form eq. ~\eqref{eq:FoMdiff}.  Let us adopt the notation  $\dot{x} \equiv \frac{dx}{d\varepsilon}$

We find from equation \ref{eq:Z} that 
\begin{eqnarray}	
   \dot{X} &=&  \Im 
   \Big \{Z_0 \dot{\zeta}
   \frac{(1-\tanh^2 (\gamma l))}{(1+\zeta \tanh(\gamma l))^2}
   \Big \}\nonumber\\
    &=& \frac{1}{\varepsilon} \Im 
    \Big \{  Z_0 \zeta
    \frac { (\tanh^2 (\gamma l)-1)}{(1+\zeta\tanh(\gamma l))^2 }
    \Big \}
    \label{eq:Zetadot}
\end{eqnarray}
 and, according to our definition
\begin{eqnarray}
    \dot{\text{FoM}}&=&
   \frac{\partial \text{FoM} (\varepsilon)}{\partial \varepsilon} = 
   \frac{|\dot{X}|}{2 R}  \nonumber\\ 
   &=&
   \frac{1}{2 \varepsilon} 
   \frac{
   \Im \Big \{ Z_0 \zeta \frac { (\tanh^2 (\gamma l)-1)}{(1+\zeta \tanh(\gamma l))^2 } \Big \}
   }
   {\Re \Big \{ Z_0
   \frac {\zeta+\tanh(\gamma l)}{1+\zeta \tanh(\gamma l)} \Big \} 
   }
   \label{eq:dFoM}
\end{eqnarray}
	
We can examine the asymptotic behavior of $\dot{\text{FoM}}$. For both asymptotically large capacitance	(or $\zeta\longrightarrow 0$), or  $\zeta\longrightarrow \infty$, clearly the integrand of the FoM also tends to zero, $\dot{\text{FoM}}\longrightarrow 0$.

We understand this as a mismatch of the impedance of the capacitor to the characteristic impedance of the transmission line.
Thus, we expect a maximum value of $\dot{\text{FoM}}$ for some intermediate value of the capacitance, about where the magnitude of the reactance of the capacitor matches the characteristic impedance of the transmission line. The exact location of this maximum can be found numerically using equation  \ref{eq:dFoM}. 

Since the heat flow through the wafer, and therefore the cooling power, is proportional to the capacitance, we will be interested in geometries with capacitance larger than the optimum for the FoM. For a given wafer size, stacking multiple wafers increases the heat flow in proportion to the number of stacked wafers reducing the equivalent capacity. This allows us to start with a large capacitance wafer, well below optimum for the FoM but with a high heat flow, and improve the FoM by stacking multiple wafers.

The objective of the tuner is to enable a large change in reactive power to be presented to a resonant cavity. Assuming we are below the high voltage breakdown limit in the tuner, the amount of reactive power that can be handled depends on two factors: The Figure of Merit and the maximum power dissipation that can be handled by the tuner. The maximum power that is allowed at the capacitor will be discussed in the next section.

\section{Physical Realisation}\label{Physical}

 The transmission line from the cavity's coupler port is terminated by a capacitor with a fast ferroelectric dielectric material. A simple exemplar configuration of the transmission line and ferroelectric wafers is shown schematically in Figure \ref{fig:Slide1}. The ferroelectric material is shown in green, copper in yellow.  The dimensions are in proportion to the example given below in Table 1.
 
 The electric connection and cooling arrangements for the inter-wafer electrode are not shown in this figure. The modulation voltage of the ferroelectric is to be applied to the spacer electrodes between pairs of ferroelectric wafers. The connection has to include a notch filter at the frequency of the cavity to avoid shorting out the RF across the capacitors. The cavity to be tuned is connected at the open end of the transmission line through an electrically coupled probe or magnetically coupled loop. Cooling of the inner conductor of the transmission line can be provided through the loop. 

The electric field required to achieve a $40\%$ change in the dielectric constant of the material is $8\,\textrm{V/}\mu\textrm{m}$ far below the breakdown electric field in the ferroelectric material of $20\,\textrm{V/}\mu\textrm{m}$. A ferroelectric thickness of $2\,\textrm{mm}$ for example would require a high voltage of $16\,\textrm{kV}$.  If the number of ferroelectric wafers is larger than two, the high voltage required for a given bias field can be reduced by a factor of two by applying alternating positive and negative bias to the inter-wafer electrodes.

The properties of the ferroelectric material that we use in this work are taken from \cite{Eucild_EIC2020}.  The ones we need are listed for convenience here.

The ferroelectric material we use is BaTiO$_3$/SrTiO$_3$-Mg:
\begin{itemize}
    \item The relative dielectric constant $\varepsilon$ is $\approx160$. At this value, $\varepsilon$ is tunable by about $40\%$ by changing the electric field from $0$ to $8\,\textrm{MV/m}$.
    \item Loss tangent $ \tan\delta \approx  3.4x10^{-4}$ at $80\,\textrm{MHz}$ \cite{FRT_techreport_2020}
    \item $\tan\delta$ scales up slowly with the frequency in the RF and UHF bands.
    \item Breakdown electric field $20\,\textrm{MV/m}$.
    \item Thermal conductivity $k=7.02\,\textrm{W/m-degree K}$
\end{itemize}

The thermal management is important for a high reactive power tuner.  Moderate cooling can be achieved by convection if gas or liquid are introduced into the capacitor section, or forced fluid circulation for highly demanding applications. In the latter case, the cooling lines should have insulated inserts to prevent shorting out the RF voltage present on the spacer electrodes.  For all cases where the ferroelectric wafers are sufficiently cooled from both sides, the maximum temperature rise in the ferroelectric material will be given by:
\begin{equation}
    \Delta T = \frac{w P_f}{2 k N_w S}
    \label{eq:deltaT}
\end{equation}

Where $P_f$ is the total power dissipated in the ferroelectric volume, $w$ is the thickness of each of the  wafers and $S$ is the surface area of each of the ferroelectric plates (including both sides of each wafer). $N_w$ is the number of wafers and  k is the thermal conductivity coefficient of the ferroelectric material with a measured value of k=7.02 Watt/meter degree K.

The design involves the choice of the capacitor’s form factor $S/w$ with the objectives of meeting the power dissipation objectives and acceptable FoM. A small capacitance is desirable in order to increase the reactance of the tuner and thus the frequency tuning range, whereas a larger capacitance increases the power dissipation capability.

At a given surface area, we can increase the capacitance and the power handling in the same proportion by decreasing the width of the ferroelectric plates.  This would increase the power handling capability but decreases the required high voltage and could make the device less robust.

As can be seen from equation \ref{eq:deltaT}, for a given maximum allowed temperature rise, scaling the surface area $S$ and the gap  $w$ by the same factor, does not change the capacitance or the power handling  $P_{f}$. 

We note that by subdividing the capacitor to m units of width $w/m$ (keeping a constant $S$), which are connected in series RF wise, but in parallel as far as the bias and cooling are concerned, we reduce the required bias voltage by a factor of $m$ and increase the cooling capacity by a factor of $m$. However, doing this increases the complexity of the unit. 

In summary, there are several advantages to this style of FE-FRT geometry compared to the previously used coaxial geometry:
\begin{enumerate}
    \item The required high-voltage is inversely proportional to the number of wafers.
    \item A high FoM and excellent heat removal can be achieved by stacking wafers.
    \item There is no bias high voltage on the transmission line or coupler.
	\item The ferroelectric material consists of flat plates, easily manufactured with precision to achieve good electric and thermal contacts.
	\item Cooling the ferroelectric is simple and effectual, allowing a very high power level to be achieved.
\end{enumerate}

\section{A basic example}\label{Basic}

To demonstrate the efficacy of the design of the FE-FRT tuner presented above we consider a specific example by carrying out parameter scans with the approximate analytic model described above and comparing the analytic results to precise numerical simulations.
To carry out this calculation, we have to make various choices. We take the frequency as 80 MHz and we choose the details of the transmission line, which is based on engineering consideration of the integration of the tuner with the cavity. We assume a reasonable value for the maximum allowed temperature in the ferroelectric material of $50\,^\circ\textrm{C}$. 

In the following we will assume a design based on the parameters given in Table I.

\begin{table}[hbt]
   \centering
   \caption{List of parameters used in the specific example of the ferroelectric tuner.}
   \begin{tabular}{lcc}
       \hline\hline
       \textbf{Parameter}     & Value        & \textbf{Units} \\
       \hline
           RF frequency           & $80$   & MHz     \\ 
           Cu conductivity $\sigma$   &$6 \cdot 10^7 $  & S/m    \\ 
           FE loss tangent           & $3.4\cdot 10^{-4}$   & --     \\ 
           Number of wafers           & 2      &-- \\ 
           Wafer radius                & 25      & mm  \\
         
           Wafer thickness            & 2.5      & mm      \\ 
           Coax. inner conductor radius $a$     & 26.9   & mm      \\ 
           Coax. outer conductor radius $b$     & 73.18   &  mm       \\
           Transmission line length          & 200       &  mm        \\
           Inter-wafer electrode thickness      & 5         & mm      \\ 
           Inter-wafer electrode radius         & 25    & mm      \\ 
       \hline\hline
   \end{tabular}
   \label{TBL_SWSP}
\end{table}

The two wafers of ferroelectric material may dissipate a total of 2.2 kW while not exceeding a temperature rise of 50 degrees C, thanks to the large area, thin width and four cooling surfaces. We find out that this particular design can cause a change in reactive power of 2.2 MVAR given the figure of merit of 158. The $\text{FoM}_C$ value is 494, which means that for each kilowatt we may dissipate in the capacitor we get nearly a megawatt reactive power for the cavity. The performance just discussed, of the FE-FRT specified in Table~\ref{TBL_SWSP}, is summarised in Table~\ref{TBL_PERF}.

\begin{table}[hbt]
   \centering
   \caption{List of parameters used in the specific example of the ferroelectric tuner.}
   \begin{tabular}{lcc}
       \hline\hline
       \textbf{Parameter}     & Value        & \textbf{Units} \\
       \hline
           Ferroelectric only FoM           & $494$   & --     \\ 
            Total FoM   &$158$  & --    \\ 
           Maximum total dissipated power           & $7$   & $\textrm{kW}$     \\ 
           Maximum dissipated power in ferroelectric           & $2.2$   & $\textrm{kW}$     \\ 
           Maximum change in reactive power    & $2.2$      & $\textrm{MVar}$ \\ 
       \hline\hline
   \end{tabular}
   \label{TBL_PERF}
\end{table}

By applying a larger number of wafers, the reactive power can be considerably increased, due to the increase in cooling area of the ferroelectric as will be demonstrated in the next section. 

To gain better insight into the dependence of the FoM on various parameters, we performed scans of selected parameters while keeping the other parameters at the values specified in Table 1.  In the following figures, we compare the analytic expressions derived above with finite-element code, using software simulations performed in CST Studio Suite\textsuperscript{\textregistered}.

\begin{figure}[htb]
\centering
\includegraphics[width=0.9\columnwidth, angle=0,origin=c]{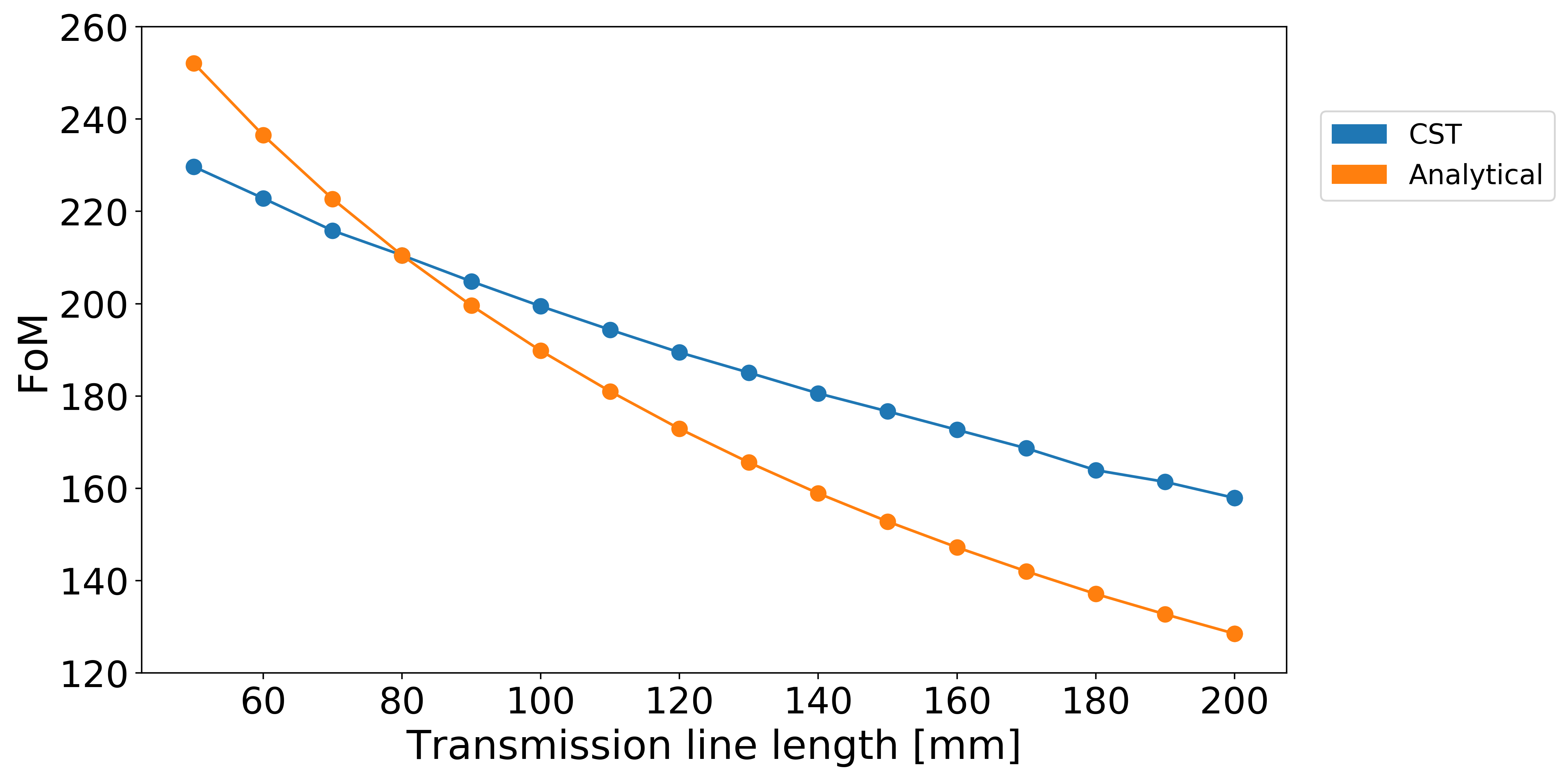}
\caption{\label{fig:Slide2} 
The Figure of Merit as a function of the length of the transmission line.}
\end{figure}

To appreciate the effects of losses in the transmission line on the general FoM, we produce Figure \ref{fig:Slide2}. While the length of the transmission line does not affect the $\text{FoM}_C$, it still affects the overall FoM and therefore should be kept short.

\begin{figure}[htb]
\centering
\includegraphics[width=0.9\columnwidth, angle=0,origin=c]{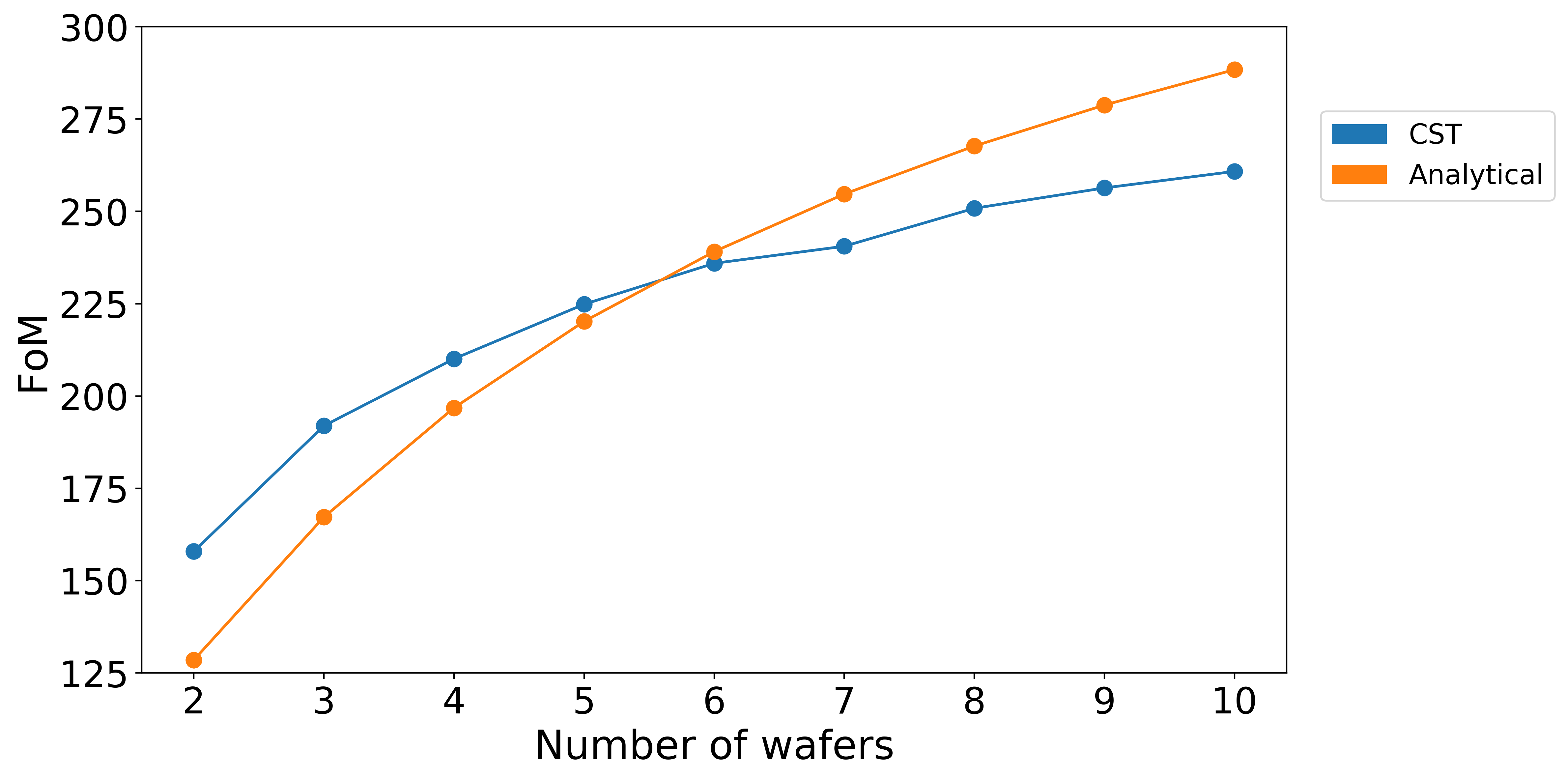}
\caption{\label{fig:Slide3} 
The Figure of Merit as a function of the number of wafers.}
\end{figure}

\begin{figure}[htb]
\centering
\includegraphics[width=0.9\columnwidth, angle=0,origin=c]{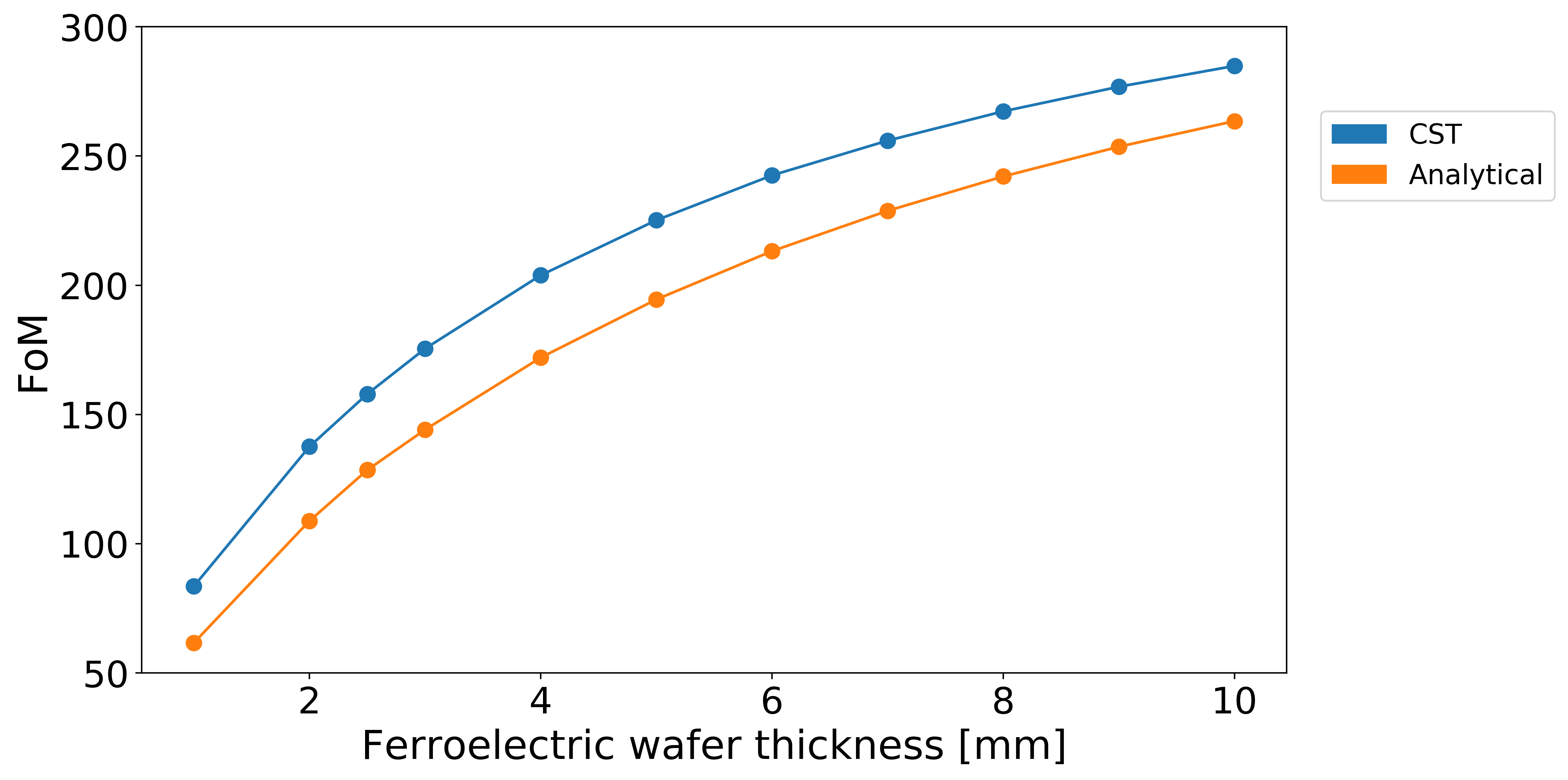}
\caption{\label{fig:Slide4} 
The Figure of Merit as a function of the wafer thickness.}
\end{figure}

The next two figures, \ref{fig:Slide3} and
\ref{fig:Slide4}, show a similar trend, with the FoM increasing as we increase the number of 2.5 mm thick wafers connected electrically in series, or the thickness of the wafers in a two-wafer configuration. 

The reason for the similarity is that in both cases we just decrease the capacitance of the tuner, leading towards the peak value of the FoM as discussed above in the analysis of equation  \ref{eq:dFoM}. The reason for the somewhat inferior FoM values in figure  \ref{fig:Slide3}, is the added RF losses in the inter-wafer electrodes. However, increasing the number of wafers improves the cooling capability and reduces the value of the high-voltage bias in direct proportion to the number of wafers connected in series at a cost of increased construction complexity. 

The reasonable agreement we get for these two calculations indicate that our analytical model is conceptually correct and a useful tool for initial design work. To understand why there are differences we must remember that for the analytic calculation of $Z$ and the $\textrm{FoM}$ we treat the ferroelectric wafers as ideal capacitors joining sections of coaxial transmission line.  The reality is more complicated, the sections of the FE-FRT containing the ferroelectric are in fact short sections of waveguide filled inhomogenously with dielectric (ferroelectric wafers in the center surrounded by vacuum). Among other things, self-inductance terms as well as reflections from the waveguide, coaxial line transitions would need to be treated more rigorously if more precision from the analytic model was required.
\section{Coupling the tuner to a cavity}\label{Coupling}

\begin{figure}[htb]
\centering
\includegraphics[width=0.9\columnwidth, angle=0,origin=c]{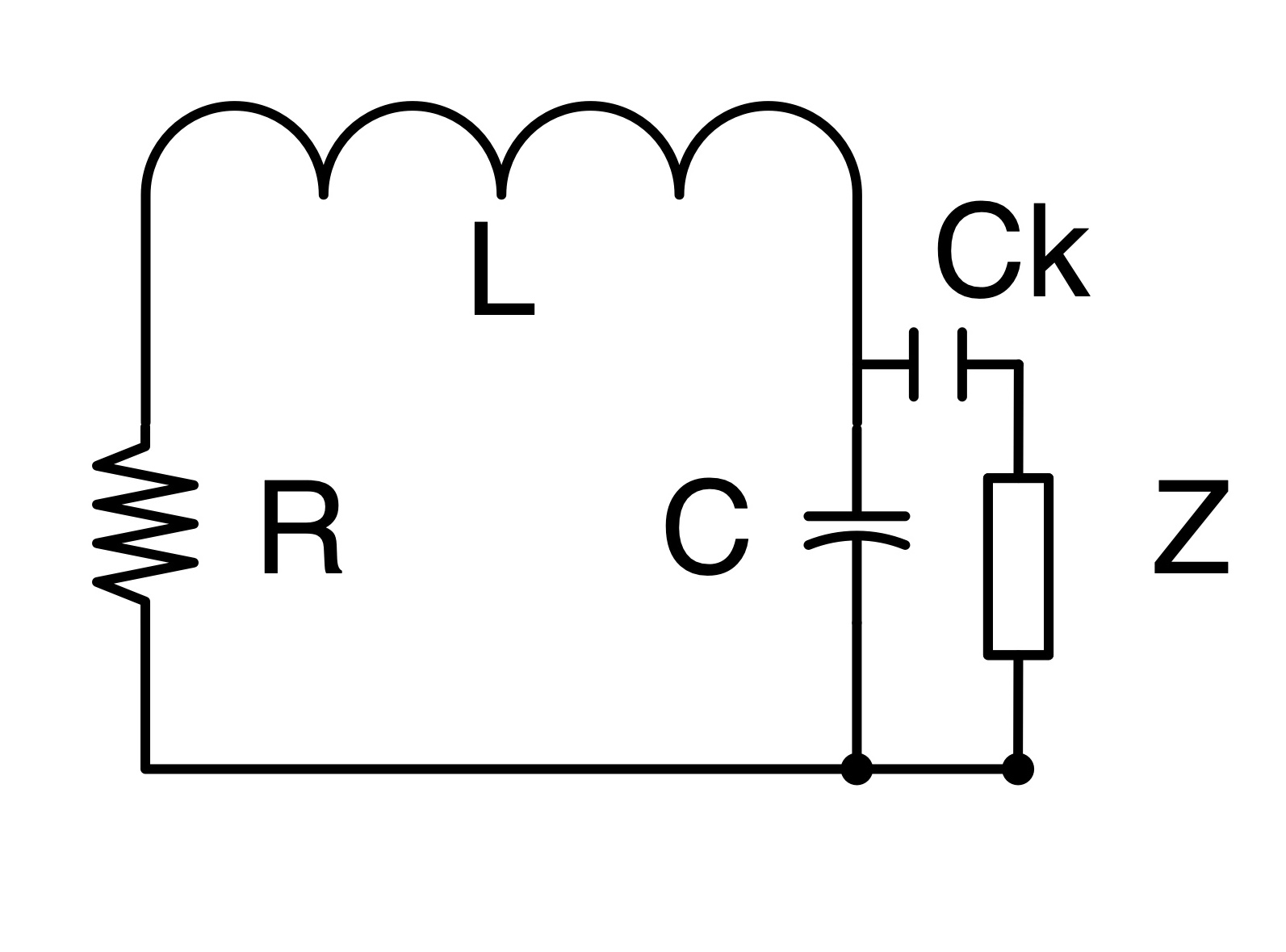}
\caption{\label{fig:Resonator} 
Equivalent circuit of a cavity coupled capacitively to a reactive tuner, represented by an impedance $Z$.}
\end{figure}

The objective of the tuner is to change the frequency of a cavity. In this section we will use a simple equivalent circuit of a cavity with an external coupling port leading to the tuner, as shown in Figure \ref{fig:Resonator}.  

In reality it is unlikely any given coupler will be well modelled by the circuit diagram shown in Figure~\ref{fig:Resonator} and therefore whilst the FoM calculated for a particular FE-FRT can be accurately modelled analytically there is no guarantee that the tuning range that can be achieved for a particular external Q will agree with the expression developed below.  The derivations below are nonetheless instructive for giving a broad insight into the behaviour that can be expected from an FE-FRT coupled to a cavity.

For really reliable predictions of the tuning range that will be achieved from a given FE-FRT coupled to a cavity with a given coupler a more complicated equivalent circuit is required to represent the coupler whose values must be found from simulation.

The cavity to be tuned is represented by $R_c$, $L_c$ and $C_c$. It has an electric coupling port through a coupling capacitance $C_K\equiv KC_c$. The load $Z=R+jX$ may represent a real characteristic impedance $Z_0$ for the determination of the external Q of the port. It may also be used to consider the frequency shift introduced by the impedance of tuner, as developed in sections \ref{FoM} and \ref{Analytic}.
 
 Let us make a couple of definitions aimed at simplifying the equations:
$\omega_0 \equiv \frac{1}{\sqrt{L_cC_c}};$  
$   Q_0 \equiv \frac {\omega L_c}{R_c}$

The impedance (normlized to $\omega L$) of the circuit in Figure \ref{fig:Resonator} is

\begin{equation}	
    \frac{Z}{\omega L} = \frac{1}{Q_0}+j(1-\frac{\omega_0^2}{\omega^2}F) 
    \label{eq:Resonant}
\end{equation}

 $F$ is a complex function of $C_K$ and $Z$, modifying the capacitance $C_c$ due to the parallel-connected load.

Then we find that $F$ is given by
\begin{eqnarray}
 F&=&\frac{R+jX+\frac{1}{j \omega C_K}}{\frac{1}{j \omega C_c} + \frac{1}{j \omega C_K} + R +jX}\nonumber\\
 &=& 
 \frac{j \omega C_c R - \omega C_c X + \frac{1}{K}}{j \omega C_c R - \omega C_c X + \frac{1}{K} +1}
\label{eq:F}
\end{eqnarray}

As a first application of equation \ref{eq:F}, let us take $R=Z_0$, $X=0$, just a real load. We inspect the imaginary part of $F$, which is responsible to adding losses to the resonator circuit.

\begin{equation}	
    \Im(F) = \frac {\omega C_c Z_0 K^2}{(\omega C_c Z_0 K)^2 + (K+1)^2}
    \label{eq:ImF}
\end{equation}
We plug equation \ref{eq:ImF} into \ref{eq:Resonant}, with the approximations $\omega_0 \approx \omega$ and $\omega C_c Z_0 K \ll 1$. We notice an extra loss term, which we identify as $1/Q_e$, where

\begin{equation}	
    \frac{1}{Q_e} \approx \frac {Z_0 K^2}{\omega L_c}
    \label{eq:Qe}
\end{equation}
We have identified K as the coupler’s transformer turn ratio. 

Now we apply equation \ref{eq:F} to evaluate the frequency tuning of the resonator induced by the real part of $F$. Setting $R=0$ in  equation \ref{eq:F}, we get

\begin{equation}	
    \Re(F) = F = \frac {\omega C_c K X -1}{\omega C_c K X  -K-1}
    \label{eq:RealF}
\end{equation}

To determine the modification of resonant frequency of the circuit, we equate to zero the imaginary term in equation \ref{eq:Resonant}:

\begin{equation}	
   \Big ( \frac{\omega_0}{\omega}\Big )^2 = \frac {\omega C_c K X -K-1}{\omega C_c K X-1}
    \label{eq:Resmod}
\end{equation}
    
    Equation \ref{eq:Resmod} provides a key relationship between the frequency of the resonator which we wish to tune and the reactance $X$ of the tuner connected to the cavity. 
Specifically, the frequency shift between the initial state and the final state of the ferroelectric can be obtained through equation \ref{eq:Deltaf}

\begin{equation}	
   \frac{\Delta \omega_{if}}{\omega} = \frac{1}{2} \Big ( \frac {\omega C_c K X_f -K-1}{\omega C_c K X_f -1} - \frac {\omega C_c K X_i -K-1}{\omega C_c K X_i -1} \Big)
    \label{eq:Deltaf}
\end{equation}
    
    The frequency shift increases with $K$ and with $X$. The magnitude of $K$ is technically limited. The magnitude of $X$ is an available parameter, with some restrictions. When we are dealing with a small frequency tuning range, such as compensating for acoustic noise in a superconducting resonator, we can make the ferroelectric capacitor small to increase $X$. For a large frequency tuning application, such as compensating for beam - cavity interactions, we need a large reactive power, which is difficult to dissipate in a small capacitor. For such applications we resort to the technique described in section \ref{Application}.
    
    Using the resonant circuit, we can also determine the reactive power incident on the tuner port in the two states, as well as the change in reactive power $\Delta\mathcal{P}_{if}$. This is useful to determine the absorbed power in the tuner elements through the FoM.
    
\begin{equation}\label{eq:ReactiveP}
     \Delta\mathcal{P}_{if}=\mathcal{P}_{f}-\mathcal{P}_{i}=
     \frac{ U \omega ^2 C_c K^2 X_f}{( 1+\omega C_c K X_f )^2} -\frac{ U \omega ^2 C_c K^2 X_i}{ ( 1+\omega C_c K X_i )^2} 
\end{equation}

Equations \ref{eq:Deltaf} and \ref{eq:ReactiveP} are closely related. If we make the approximation $\omega C_c K X \ll 1$ (meaning that the load impedance is much smaller that the impedance of the coupling capacitor) then they combine to yield the familiar

\begin{equation}\label{eq:PeqUdOmega}
     \Delta\mathcal{P}_{if}=2 U \Delta \omega_{if} 
\end{equation}

We observe that if the signs of $X_i$ and $X_f$ were opposed, the change in reactive power will be greater than either incident reactive power terms. One way this can be achieved whilst keeping the transmission line short to minimise losses would be by connecting an inductance in parallel to the variable capacitor, as will be shown in section \ref{Application}. The value of the inductance can be chosen such that the values of the reactances $X_i$ and $X_f$ as seen by the cavity are equal and opposite.

Finally, we wish to choose correctly the external coupling $Q_e$. We still use the approximation $\omega C_c K X \ll 1$. 
Combining equations \ref{eq:ReactiveP}, \ref{eq:PeqUdOmega} and $\ref{eq:Qe}$, and we get the required $Q_e$ for the tuner port (as stated this assumes the cavity and coupler are well modelled by Figure~\ref{fig:Resonator} which may not be the case in reality):

\begin{equation}\label{eq:Qefinal}
     Q_e= \frac{1}{2}\frac{\omega}{\Delta \omega_{if}} \frac{\Delta X_{if}}{Z_0}
\end{equation}

Now we are in a position to apply some simple scaling laws for a very simple case of a cavity tuned by a variable capacitor. From equation  \ref{eq:deltaT} we get 

\begin{equation}\label{eq:PropP}
     P\propto\frac{N_w S}{w}
\end{equation}

The impedance connected to the cavity is given by \ref{eq:Zcapacitor}. Therefore we get the proportionality

\begin{equation}\label{eq:PropZ}
     \Delta X_{if}\propto\frac{N_w^2 }{P}
\end{equation}

We learn from equation \ref{eq:Qefinal} that the product of $Q_e$ and the tuning range is proportional to $\Delta X_{if}$. Then the significance of equation \ref{eq:PropZ} is clear: For a given dissipated power $P$, the product of the tuning range and $Q_e$ scales as the number of wafers squared.

For a low power dissipation or small tuning range, a small number of wafers suffices. For demanding applications, combining a lower limit of $Q_e$, a large tuning range and high power, we may rely on enhancement of the impedance by the transmission line.

\section{Application to a realistic, ultra-high reactive-power case}\label{Application}
There are numerous applications of fast, high reactive-power tuners for particle accelerators. Aiming to demonstrate the feasibility of an ambitious goal, we apply the FE-FRT technique to one specific RF cavity out of the many in the CERN accelerator complex.

The CERN PS accelerator has three single-cell RF-cavities with a resonance frequency of 80 MHz.  These cavities are currently equipped with mechanical tuners for frequency shifting to accommodate the acceleration of protons and ions.

We describe the application of an FE-FRT tuner to this RF cavity. Our objective is to demonstrate an ambitious tuning range of 230 kHz of the fundamental mode, at a cavity stored energy of 3.2 Joule, which translates to approximately 10 MVAR differential reactive power.

\begin{figure}[htb]
\centering
\includegraphics[width=0.9\columnwidth, angle=0,origin=c]{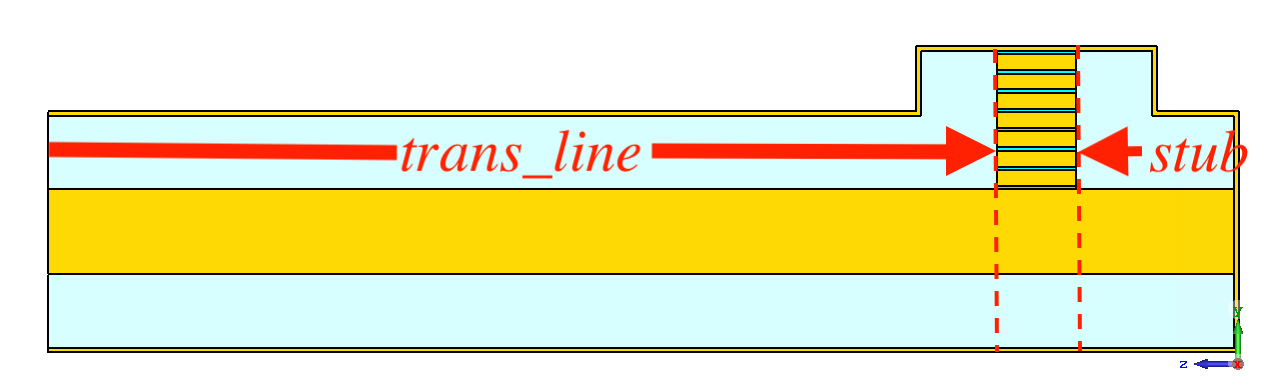}
\caption{\label{fig:Tuner_PS} 
The parallel stub tuner, with a single wafer-stack. The tuner comprises a ferroelectric wafer stack in parallel to the "stub", and a transmission line leading to the cavity.}
\end{figure}

The proposed tuner is shown in Figure \ref{fig:Tuner_PS} and it's dimensions are given in Table II. The short transmission line in parallel to the capacitor stack, the stub, serves as an inductance as discussed in the previous section.
   
The length of the stub is adjusted such that $X_f=-X_i$. The reactance of the device as seen by the cavity (in either state) is significantly higher than the reactance of the capacitor. This feature allows a higher $Q_e$ and larger tuning range, as well as a lower peak power in the coupler. Therefore this configuration is ideal for ultra-high reactive power applications involving fast binary switching between the two states of the ferroelectric. For less demanding applications, proportional control can also be achieved with $X_i$ and $X_f$ both positive and using the configuration depicted in $\ref{fig:Slide1}$.
   
Our design provides a high FoM, high reactive power tuning and good impedance for coupling to the cavity. 

The cooling of the inner conductor of the transmission can be done from the shorted end. Either electric coupling or magnetic coupling are possible. As an additional option the single  stack of wafers can be replaced by a few stacks connected in parallel.  This last feature would allow pulsing the bias voltage on different stacks with timings, enabling piece-wise arbitrary reactive power functions to be achieved with highly efficient pulsed-power modulators.  

We implemented the design in the analytic model (not shown here for the sake of brevity) and searched for an acceptable solution, finding the set of parameters given in Table~\ref{TBL_PSFRT} is satisfactory.  We expect that the transmission line can be easily cooled by circulating a coolant from the shorted end of the line. Therefore we use the expression:
\begin{equation}
    2\textrm{FoM}_CP_f
\end{equation}
to determine the maximum change in reactive power which is given in Table~\ref{TBL_PSFRT}.

As mentioned previously however the tuning range and external Q have been calculated assuming the coupler is correctly modelled as shown in Figure~\ref{fig:Resonator}.  In future work a more complete equivlant circuit model to describe the coupler will be derived from simulations and some of the dimensions shown in Table~\ref{TBL_PSFRT} can be expected to change.  The final design will and tuning range will also be verified with simulation.  Crucially however the FoM and maximum change in reactive power can be expected to be be very similar to what is shown.

\begin{table}[hbt]
   \centering
   \caption{List of parameters for the PS cavity ferroelectric tuner.}
   \begin{tabular}{lcc}
       \hline\hline
       \textbf{Parameter}     & \textbf{Value}        & \textbf{Units} \\
       \hline
           Number of wafers           & $8$   &--      \\ 
           Wafer diameter $d$   &$48 $  & mm    \\ 
           Wafer thickness           & $2$   & mm     \\ 
           Coax outer conductor radius $b$           & 73.18      & mm \\ 
           Coax inner conductor radius $a$             & 26.9      & mm  \\
         
           Stub length  $L$         & 105     & mm      \\ 
           Transmission line length $l$     & 60   &  mm       \\
           Ferroelectric absorbed power limit          & 10.1       &  kW        \\
           Capacitor's $\textrm{FoM}_C$      & 497         & --      \\ 
           System FoM        & 282      &  --        \\
                Change in reactive power        & 10    & MVAR     \\ 
           Frequency tuning         & 230       &  kHz       \\
       \hline\hline
   \end{tabular}
   \label{TBL_PSFRT}
\end{table}

\section{Summary}\label{Summary}

We presented a novel design of a ferroelectric fast reactive tuner comprising a short coaxial transmission line terminated by a capacitor formed from multiple ferroelectric wafers connected to the high voltage in parallel but seen by the RF in series. The simple geometry of the wafers allows easy fabrication and efficient cooling whilst the use of multiple wafers further increases the power handling and reduces the required high voltage at the same time achieving a very high figure of merit.

We presented an analytical model backed up by finite element numerical simulation of the device. We studied the parametric dependence of the tuner, and have shown its capability to provide in excess of a Mega-VAR of reactive power tuning at high figure of merit values. 

Another novel design of an FE-FRT tuner was introduced for the PS-cavity containing a stub.  This enables a higher change in reactive power to be achieved with reduced peak ractive power.  In addition it is easily possible to include multiple capacitor stacks in such a tuner allowing a shaped temporal profile of the modulation using efficient pulsed-power devices and an additional degree of freedom in the design.

\begin{acknowledgments}
We gratefully acknowledge the support of our colleagues on related aspects of our research on ferroelectric tuners, A. Kanareykin, C. Jing (Euclid Techlabs LLC) and G. Burt (Lancaster Univ.). One of the authors (IBZ) acknowledges work at BNL under the US Department of Energy Office of Science contract DE-SC0012704.
\end{acknowledgments}

\end{document}